\documentclass{aa}
\usepackage{graphicx}
\usepackage{natbib}
\usepackage{txfonts}
\usepackage{color}
\begin{document}
    \title{HD 50844: the new look of $\delta$ Sct stars from CoRoT space
photometry \thanks{
The CoRoT space mission was developed and  is operated by the French
space agency CNES, with participation of ESA's RSSD and Science Programmes,
Austria, Belgium, Brazil, Germany, and Spain.
This work is based on ground--based observations made with ESO telescopes
at the La Silla Observatory under the ESO Large Programme LP178.D-0361
and on data collected at the Observatorio de Sierra Nevada (Spain), at
the Observatorio Astron\'omico Nacional San Pedro M\'artir (Mexico), and at
the Piszk\'estet\"o Mountain Station of Konkoly Observatory 
(Hungary).}$^,$\thanks{Table 2 is only available in electronic form
at the CDS via anonymous ftp to cdsarc.u-strasbg.fr (130.79.128.5)
or via {\tt http://cdsweb.u-strasbg.fr/cgi-bin/qcat?J/A+A/}}
}

   \author{
E.~Poretti\inst{1}, 
E.~Michel\inst{2},
R.~Garrido\inst{3},
L.~Lef\`evre\inst{2},
L.~Mantegazza\inst{1},
M.~Rainer\inst{1},
E.~Rodr\'{\i}guez\inst{3},
K.~Uytterhoeven\inst{1}
\thanks{Current address: Laboratoire AIM, CEA/DSM CNRS
Universit\'e Paris Diderot; CEA, IRFU, SAp, centre de Saclay, 91191,
Gif-sur-Yvette, France},
P.J.~Amado\inst{3},
S.~Mart\'{\i}n-Ruiz\inst{3},
A.~Moya\inst{3},
E.~Niemczura\inst{4},
J.C.~Su\'arez\inst{3},
W.~Zima\inst{5},
A.~Baglin\inst{2},
M.~Auvergne\inst{2},
F.~Baudin\inst{6},
C.~Catala\inst{2},
R.~Samadi\inst{2},
M.~Alvarez\inst{7},
P.~Mathias\inst{8},
M.~Papar\`o\inst{9},
P.~P\'apics\inst{9},
\and
E.~Plachy\inst{9}
}
\authorrunning{Poretti et al.}
\offprints{E.~Poretti\\
\email{ennio.poretti@brera.inaf.it} }

   \institute{INAF -- Osservatorio Astronomico di Brera,
              Via E. Bianchi 46, 23807 Merate (LC), Italy
              \and
LESIA, Observatoire de Paris, CNRS (UMR 8109), Universit\'e Paris 6,
    Universit\'e Paris Diderot,  
              5 pl. J. Janssen, 92195 Meudon, France
\and
Instituto de Astrof\'{\i}sica de Andaluc\'{\i}a, Apartado 3004, 18080 Granada, Spain
\and
Astronomical Institute of the Wroclaw University, ul. Kopenika 11, 51-622 Wroclaw, Poland
\and
Instituut voor Sterrenkunde, K.U.~Leuven, Celestijnenlaan 200 D, 3001 Leuven, Belgium
\and
Institut d'Astrophysique Spatiale, CNRS, Universit\'e Paris XI UMR 8617, 91405 Orsay, France
\and
Observatorio Astron\'omico Nacional, Instituto de Astronomia, UNAM, Apto Postal 877,
Ensenada, BC 22860, M\'exico
\and 
UMR 6525 H. Fizeau, UNS, CNRS, OCA, Campus Valrose, F-06108 Nice Cedex 2, France
\and 
Konkoly Observatory, PO Box 67, 1525 Budapest, Hungary
             }

   \date{Received, accepted}
   \abstract
   {The ground--based campaigns on  $\delta$ Sct stars revealed richer and richer frequency spectra of
these opacity--driven pulsators,
 thanks to  continuous improvements in their exploitation. 
It has also been suggested that the detection of a wealth of very low amplitude 
modes was only a matter of signal--to--noise ratio.
Access to this treasure, impossible from the ground, is one of the scientific 
aims of the space mission CoRoT.
}
{This work presents the results obtained on HD~50844, the only $\delta$ Sct star observed in the CoRoT Initial
Run (57.6~d). 
The aim of these CoRoT observations was to investigate and characterize for the first time the pulsational behaviour 
of a $\delta$ Sct star, when observed at a level of precision and with a much better duty cycle than from the ground.}
 {The 140,016 datapoints  were analysed  using independent approaches
(SigSpec software and different iterative sine-wave fittings) and several checks 
performed (splitting of the timeseries in different subsets, investigation of the residual light curves and spectra).
A level of 10$^{-5}$~mag  was  reached in the amplitude spectra of the CoRoT
timeseries. The space monitoring was complemented by ground--based high--resolution spectroscopy,
which allowed the mode identification of 30~terms.   
}
   {The frequency analysis of the CoRoT timeseries  revealed hundreds of terms in
the  frequency range 0--30~d$^{-1}$. All the cross--checks  confirmed this new result. 
The initial guess that $\delta$ Sct stars have a  very rich frequency content
is confirmed. The spectroscopic mode identification gives  theoretical support since 
very high--degree modes (up to $\ell$=14)  are identified. 
We also prove that cancellation effects
are not sufficient in removing the flux variations associated to these modes at the 
noise level of the CoRoT measurements.
The ground--based observations indicate that HD~50844 is an evolved star that is  slightly underabundant 
in heavy elements, located on the Terminal Age Main 
Sequence. Probably due to this unfavourable evolutionary status, 
no clear regular distribution is observed in the frequency set. 
The predominant term ($f_1$=6.92~d$^{-1}$) has been identified as the  fundamental radial mode 
combining ground-based photometric and  spectroscopic data.
}
 {The CoRoT scientific programme contains other $\delta$ Sct stars,  with different 
evolutionary statuses. 
The very rich and dense frequency spectrum
discovered in the light curve of HD~50844 is the starting point
for  future investigations.}


    \maketitle
%

\section{Introduction}
A huge effort has been made in the past decades to observe  $\delta$ Sct stars 
from the ground. These opacity--driven pulsators show many excited
modes with amplitudes that can be reached by photometry from the ground. After several
results obtained by different teams performing single--site observations 
\citep{porvie}, the
most promising targets have been monitored by means of multisite campaigns,
mainly in the framework of the Delta Scuti Network \citep{brevie} and of
the STEPHI network \citep{stephi}. The most observed $\delta$
Sct star from the ground is FG Vir: 75 frequencies have been detected with 
amplitudes down to  0.2~mmag thanks to a multisite monitoring spanning several years \citep{fgvir}.  

The first $\delta$ Sct star monitored from space was $\theta^2$ Tauri, 
with the WIRE satellite \citep{wire}.
WIRE data, collected in only 19~days,  increased the number of the known frequencies
from 5 to 12.
More recently, the Canadian Space Agency mission MOST 
detected in the light curve of the $\delta$ Sct star HD 209775 \citep{jaymie}  
roughly the same number of frequencies as in  FG Vir.
The MOST observations,
concentrated in two runs lasting 14 and 44~days and separated by about
1~year, gives a taste of the new horizons that can be opened by a space mission, which monitors
$\delta$ Sct stars in a continuous way for several months at a time.

The asteroseismologic scientific case of the CoRoT (COnvection, ROtation and planetary Transits) space mission 
gives such an opportunity for  long--term monitoring
of the same field \citep{esa3, esa1, esa2}. \cite{flight} describe the CoRoT in--flight performances.
Some preparatory work has been  undertaken to identify potentially interesting 
targets in the selected observing zones and, among them, several $\delta$ Scuti stars have
been proposed \citep{center,anti}. After the successful launch in December
2006 and the in-orbit calibrations, CoRoT observed its first field in February
and March, 2007, i.e., the first Initial Run (IR01).
One of the new $\delta$ Sct stars 
discovered in the preparatory work, HD~50844, was among the 10 asteroseismologic targets of the IR01. 
It is a slightly evolved star, close
to the Terminal Age Main Sequence. The physical parameters derived from Str\"omgren 
photometry are $M_V$=1.31, T$_{\rm eff}$=7500~K, $\log g$=3.6, and [Fe/H]=--0.4~dex
\citep{anti}. In this paper we 
review our understanding of  $\delta$ Sct stars and unveil a new ``look" 
of their frequency spectra, thanks to the intensive CoRoT monitoring. We also try to  
answer some open questions on mode excitation in the pulsators located in the lower 
part of the instability strip.
To give as  complete as possible answers, we also obtained high-resolution
 spectroscopy and multicolour photometry from ground. In
particular, the study   of line profile variations is a complementary approach, and essential for
 the identification of the excited modes. 

\begin{figure}[]
\begin{center}
\includegraphics[width=\columnwidth,height=\columnwidth]{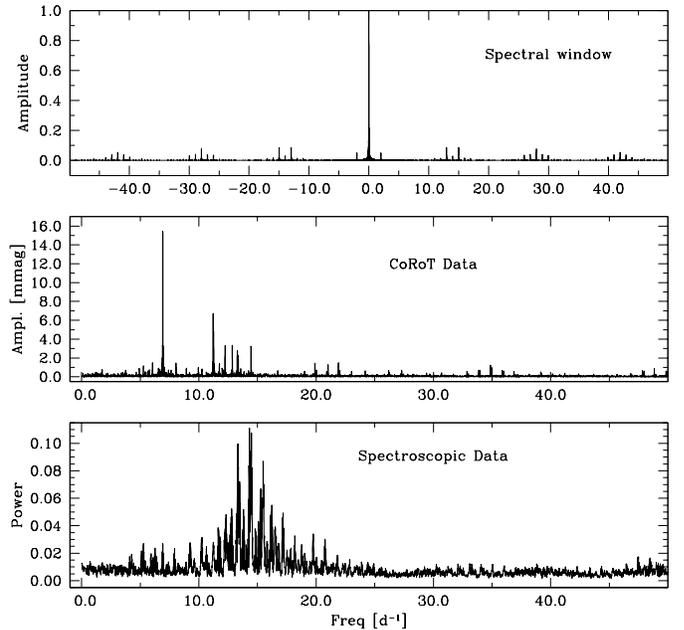}
\caption{\footnotesize {\it Top panel}:
the spectral window of the CoRoT data on HD~50844, observed in the Initial Run
of the mission. {\it Middle panel}: the amplitude  spectrum of the original CoRoT data.
 {\it Bottom panel}: the power spectrum (in units of the normalized  spectral power) of the 
line profile variations detected in the ground--based spectroscopic data.
 }
\label{win}
\end{center}
\end{figure}

\begin{figure*}[]
\begin{center}
\includegraphics[width=2.0\columnwidth,height=\columnwidth]{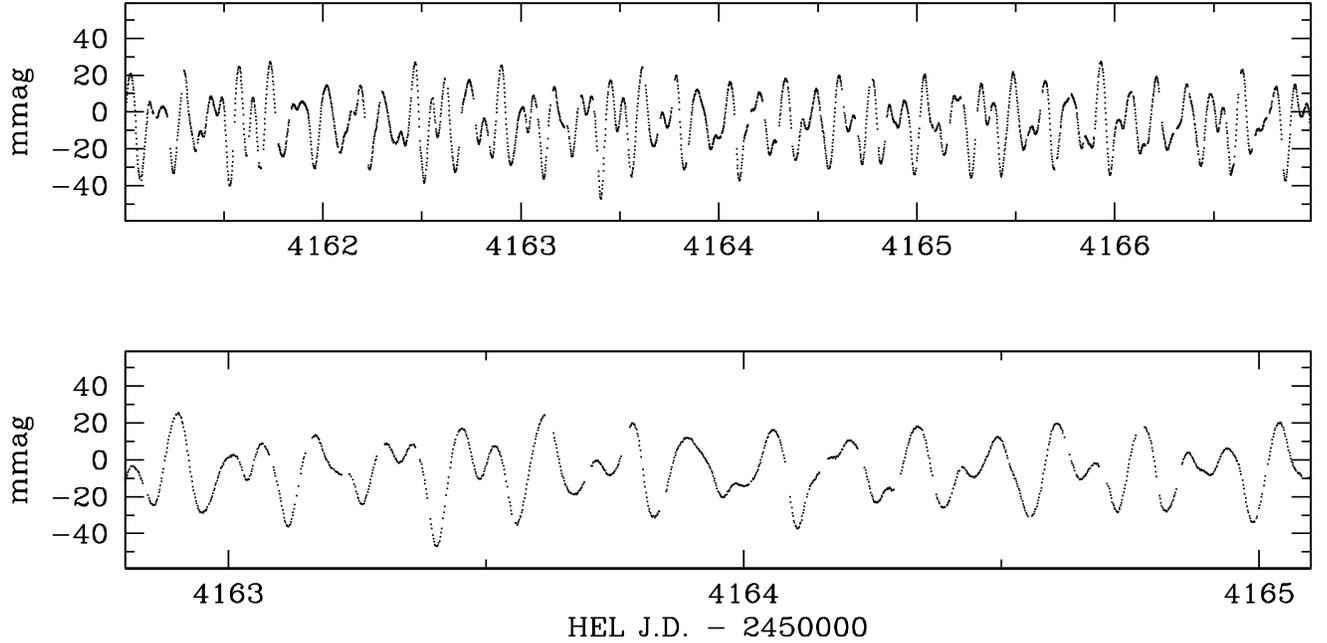} 
\caption{\footnotesize
Examples of the light curve of HD 50844 observed by CoRoT. The lower panel
is a zoom--in of a region of the upper panel. 
 }
\label{lc}
\end{center}
\end{figure*}

\section{The CoRoT data}
The IR01 started on February 2$^{\rm nd}$, 2007 and finished on March 31$^{\rm st}$, 2007   
($\Delta$T=57.61~d). The exposure time in the CoRoT asteroseismology channel is 1~sec.
For the analysis, we used the reduced N2 data rebinned at 32~sec
and  we only
considered the 140,016 datapoints for which no problem (i.e., flag=0) were reported.
The amplitudes of the background orbital variations are kept at
a  very small level by the great effectiveness of the baffle \citep{flight}. 
The subsequent rejection of the
uncertain points strongly minimized the orbital effects. 
The light curve  was detrended with a linear fit to remove the effect of ageing 
\citep{flight}.
The 32--sec time binning provided an oversampling of the light curve coverage.
Therefore,
to gain in CPU time and to reduce the noise level, we grouped the original data
into new bins of four  consecutive measurements, thus
obtaining 32,433 datapoints.

The spectral window is shown in Fig.~\ref{win}.
The rejection of the flagged points resulted in a slight
enhancement of the amplitude  of the orbital frequency, since the bad measurements
occurred mostly when the satellite  crossed the South Atlantic Anomaly (SAA). This
crossing occurred twice in a sidereal day and gave rise to an additional alias peak.
In the 0--50~d$^{-1}$ interval, there are four structures in the spectral
window. The first is composed
of a peak at 2.006~d$^{-1}$ (4\% of the amplitude) and of a lower one at 4.007~d$^{-1}$ (0.7\% of the amplitude).
The second and most prominent structure is around the orbital frequency $f_s$=13.972~d$^{-1}$:
$f_s$ (3.0\% of the amplitude), $f_s$+1~d$^{-1}$ (7.5\%) and $f_s$--1~d$^{-1}$ (8.5\%). Other structures occur
around 2$f_s$ (7.0\% of the amplitude), flanked by lower peaks at 2$f_s\pm 1$~d$^{-1}$ and 2$f_s\pm 2$~d$^{-1}$,
and around 3$f_s$ (5.0\% of the amplitude), flanked by lower peaks at 3$f_s\pm 1$~d$^{-1}$ and
3$f_s\pm 2$~d$^{-1}$.  

Several approaches were used to process the CoRoT data, e.g.,  Period04 \citep{period}, iterative
sine-wave fitting \citep{vani} and SigSpec \citep{sigspec}. 
Moreover, we refined the classical fitting proposed by \citet{vani}
 for the analysis of the CoRoT data on classical pulsators with  a
Levenberg-Marquardt algorithm.
 It uses the frequency domain as a first guess for the fitting
of sinusoids in the time domain \citep{laure}. 

Although small differences in the values
of the detected frequencies have been found (see Sect.~\ref{corot}), the main results described here are independent 
from the approach used.
The presence of the predominant term suggested in the preparatory work, based only
on ground-based 
data \citep[$f_1$=6.92~d$^{-1}$, see Fig.~5 in][]{anti}, 
has been immediately confirmed. The alias structures  of this term (e.g., $f_1+f_s$=20.89, $f_1+2f_s$=34.86, and 
$f_1+3f_s$=48.83~d$^{-1}$)
are visible in the amplitude spectrum of the original data (Fig.~\ref{win}, middle panel).
Both  methods were very effective in removing the alias structures
together with the real peak; actually, no relevant term is found at the  
alias frequencies of $f_1$. 
Taking the excellent quality of the N2 data into account, 
there is little influence of the satellite's orbital period on  the detected 
frequencies; in particular, those specifically discussed in the paper are free of 
any aliasing or orbital effect.

\section{The frequency content of the CoRoT data}
\begin{figure}[]
\begin{center}
\includegraphics[width=\columnwidth,height=\columnwidth]{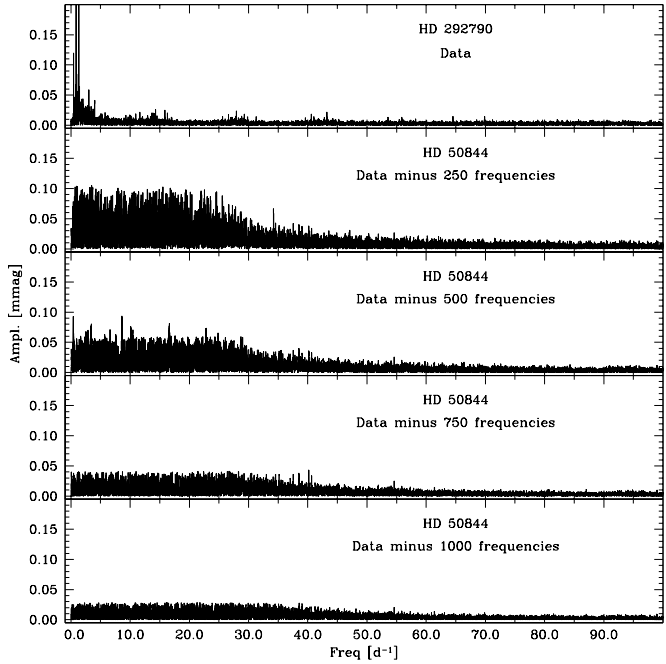}
\caption{\footnotesize
For comparison purposes, the top panel  shows the amplitude spectrum of the rotational variable
HD~292790. This star has been observed by CoRoT simultaneously
to HD~50844. The amplitude spectra of HD~50844 shown
in the other panels indicate the extreme richness of the pulsational spectrum
of this $\delta$ Sct star: up to 1000  peaks  are needed to make the  residual 
spectrum of HD~50844 comparable to the original one of HD~292790.
}
\label{spectrum}
\end{center}
\end{figure}

To have a reference frame for interpreting of the results for HD~50844 (A2, $V$=9.09), 
we also considered the data of HD~292790 (F8, $V$=9.48), which  was observed by CoRoT  in the same IR01.
Our frequency analysis shows that the luminosity of HD~292790 is modulated by the rotation in
a  simple way.
The amplitude spectrum shows the rotational frequency,
its harmonics and the satellite frequencies (Fig.~\ref{spectrum}, top panel). 
For most part of the  spectrum, i.e., from 10 to 100~d$^{-1}$, the
noise level is distributed in an uniform way and is very  low, namely below 0.01~mmag. 
For $f<5~$d$^{-1}$, where the modulation terms and the long term
drift are concentrated, the noise level increases slightly increases.
\label{corot}

In contrast,
the amplitude spectrum of HD~50844 appears to be very dense (Fig.~\ref{win}, 
middle panel, and, in a more striking way, 
Fig.~\ref{spectrum}, second panel).
 At first glance, it is clear that the amplitude spectrum of 
HD~50844 is only for $f>50~$d$^{-1}$ as flat as that of HD~292790. 
We can calculate a noise  amplitude of $7.5\times10^{-6}$ mag. 
It is straightforward to deduce that the signal is concentrated in the 
$f<30~$d$^{-1}$ domain, though there is no predominant peak 
region after identification of  250 frequencies (residual rms 1.2 mmag).
At that point, the  average peak height (0.10~mmag) for $f<30~$d$^{-1}$ is still 10  
times higher than for $f>50~$d$^{-1}$.
A huge number of terms is 
necessary to reduce the amplitude spectrum of HD~50844 at the expected noise level.  
The peak height gets  progressively lower after the identification of 
500, 750, and 1000 terms (third, fourth,  and fifth panels in Fig.~\ref{spectrum}, respectively;
see also the second column of Table~\ref{lsq}), but
the residual rms decreases very slowly (third column of Table~\ref{lsq}).
Again, many low--amplitude  peaks are necessary to 
reduce the residual rms: after detection of 500 frequencies, other 500 frequencies
with amplitudes between 0.06 and 0.03 mmag are needed to reduce the residual rms
by a factor of 0.60\%. 

\begin{table}
\begin{flushleft}
\caption{Summary of the frequency detection. 
}
\begin{tabular}{rcc rr}
\hline \hline
\noalign{\smallskip}
\multicolumn{1}{c}{Detected} & \multicolumn{1}{c}{Amplitude}&
\multicolumn{1}{c}{Residual rms} & \multicolumn{1}{c}{SNR} & \multicolumn{1}{c}{sig} \\
\multicolumn{1}{c}{frequencies}& \multicolumn{2}{c}{[mmag]}\\
\noalign{\smallskip}
\hline
\noalign{\smallskip}
0    &  \multicolumn{1}{c}{--}    & 13.63 & \multicolumn{1}{c}{--}    & \multicolumn{1}{c}{--}    \\
1    & 15.45  & 8.14  & 2060.5 &  4526.4      \\
20   &  0.64  & 2.85  &   85.5 &  175.6  \\
60   &  0.28  & 2.03  &   37.4 &   66.3  \\
100  &  0.20  & 1.74 &    26.7 &  46.9   \\
200  &  0.12  & 1.34 &    16.0 &  29.7 \\
300  &  0.09  & 1.11 &    12.0 &  23.4 \\
500  &  0.06  & 0.83 &     8.0 &  18.3 \\
750  &  0.04  & 0.61 &     5.3 &  15.7 \\
1000  & 0.03  & 0.48 &     4.0 & 12.6 \\
\noalign{\smallskip}
\hline
\label{lsq}
\end{tabular}
\end{flushleft}
\end{table}


\begin{figure}[]
\begin{center}
\includegraphics[width=\columnwidth,height=\columnwidth]{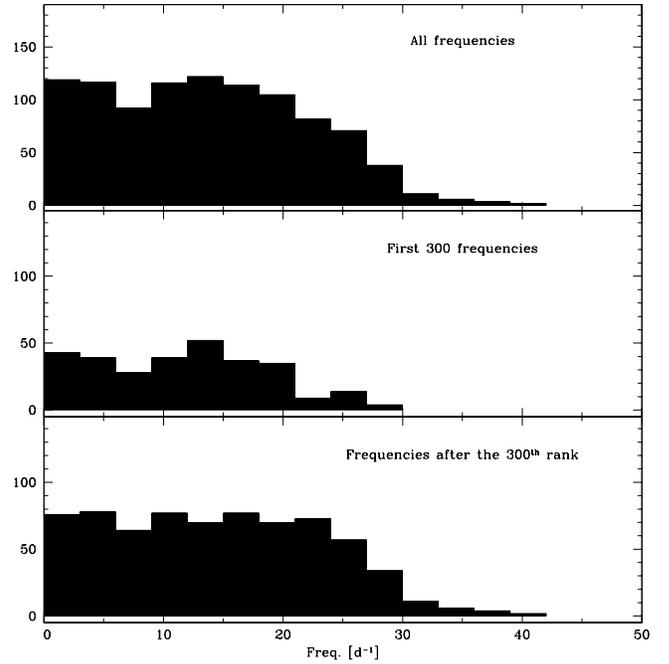}
\caption{\footnotesize Distribution of the frequencies.
{\it Top panel:} histogram of all the 1000 frequencies.
{\it Middle panel:} histogram of the first 300 frequencies in order of
detection.
{\it Bottom panel:} histogram of the frequencies detected after the first 300. 
}
\label{isto}
\end{center}
\end{figure}

The distribution of the frequencies offers some clues about the
nature of the signal. We calculated the distribution of all the frequencies,
of the first 300 in order of detection and of those detected after the
300$^{\rm th}$ rank (Fig.~\ref{isto}). The distribution of all the frequencies
is almost flat in the 0--20~d$^{-1}$ region, followed by a slow decline
until 30~d$^{-1}$.  This tail is absent in
the distribution of the first 300 frequencies, and there is a sharp step at
20~d$^{-1}$. The 20--30~d$^{-1}$ region is populated by the frequencies detected
after the first 300.

\begin{figure}[]
\begin{center}
\includegraphics[width=\columnwidth,height=\columnwidth]{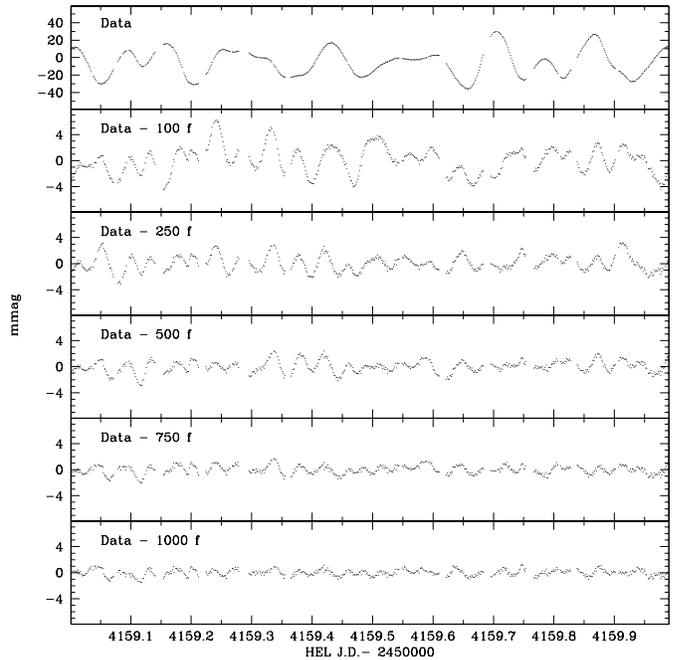}
\caption{\footnotesize
The behaviour of the light curve of HD~50844 in the original data (top panel)
and after subtracting the detected terms at different steps, where the 
time axis spans 1~d.
}
\label{residui}
\end{center}
\end{figure}
It is worthwhile investigating how the detection of so many frequencies
modifies the light curves.
Figure~\ref{residui} shows the light curve of HD~50844 at different stages of prewhitening.
The multiperiodic nature of the original data (first panel) is obvious. 
For instance, the sine wave of the predominant frequency  is almost cancelled by
interference with other terms  between HJD~2454159.5 and 2454159.6. Variability
is still evident after removing 100 frequencies. This is an important step forward 
compared to previous ground--based and space observations. Also long--term oscillations
are noticed at this level. After removing 250 frequencies, the residual light curve shows
oscillations whose periods are around 0.05~d ($\sim$~20~d$^{-1}$; third panel). After 
removing even more frequencies, the oscillations
progressively become more rapid and periods about 0.03~d ($\sim$~33~d$^{-1}$) are visible in the fourth and fifth
panels of Fig.~\ref{residui}.  This is another reflection of the distribution tail observed in the
third panel of Fig.~\ref{isto}. It is quite evident that oscillations are still
present after subtraction of 1000 terms.
As shown in the last two columns of Table~\ref{lsq}, after 1000 frequencies the SNR
is around 4.0 and the Spectral Significance  parameter (``sig") of the SigSpec method is around 12.6. 
The former is considered the acceptance limit in the frequency search algorithms such
as Period04, while
the latter is still above the threshold of the SigSpec method, i.e., 5.46. In a conservative way,
we decided to stop the frequency search at this point, as 
it becomes more and more difficult to resolve new terms within our frequency resolution.

We considered it important to check  the reliability of the frequency values determined until here.
To do so, we compared the results from SigSpec and from the iterative--sine--wave fitting,
considering an error range of 0.005~d$^{-1}$  in frequency (i.e., a width 3.4~times narrower
than 1/$\Delta$T).  
The cross--checks show an almost complete agreement on the values of the first 500 frequencies:
498  frequencies were found by both methods, whose 438 (88\%)  have an amplitude within the 20\%
range of tolerance. In particular, 
both methods correctly identified the first 462  frequencies within an interval of few units of 
$10^{-4}$~d$^{-1}$. These frequencies are listed in Table~2, which is available in electronic
form at the CDS. 
The solutions of the two methods slowly diverge in the 500--1000$^{\rm th}$ range because the 
differences will accumulate in the iterative process of the frequency extraction.
Notwithstandining, the agreement is still satisfactory when considering all the 1000 frequencies: 953 were found by both
methods, whose 805  with an amplitude within the 20\% range.   

We also performed another independent check. We created two subsets by considering   
the first and the second halves of the CoRoT timeseries.   Again, we found a strong match 
between the frequencies obtained from the two subsets. However, we
had to take  the degradation of the frequency resolution into account and had to relax the error range. 
Therefore, the same analysis was repeated considering two subsets spanning the same time baseline,
i.e., 50~days.
The first subset was composed of the data in the time intervals 0.0--10.0~d and 30--50~d 
(0.0~d is the time of the first measurement) and the second subset consisted  of the data in the 
intervals 10--30~d and 50--58~d. The two subsets have 823, 756, and 501 
frequencies (out of 1000) in common considering a maximum separation of 0.025~d$^{-1}$, 
0.020~d$^{-1}$ and 0.010~d$^{-1}$, respectively.
Such large fractions of coincident terms suggest that they are 
more likely intrinsic to the star rather than due
to transient instrumental effects.  It is possible that 
some of the lower peaks originate from changes in amplitude or phase, or  from other 
effects intrinsic to the star. In any case, 
the main conclusion of our analysis and related checks is that hundreds of terms
--~at least up to 1000~-- 
are needed  to explain the light variability of HD~50844. 
The detection of so many independent terms in the light variability of a
 $\delta$ Sct star is a totally new result.  

\begin{figure}[]
\begin{center}
\includegraphics[angle=90,width=\columnwidth,height=0.75\columnwidth]{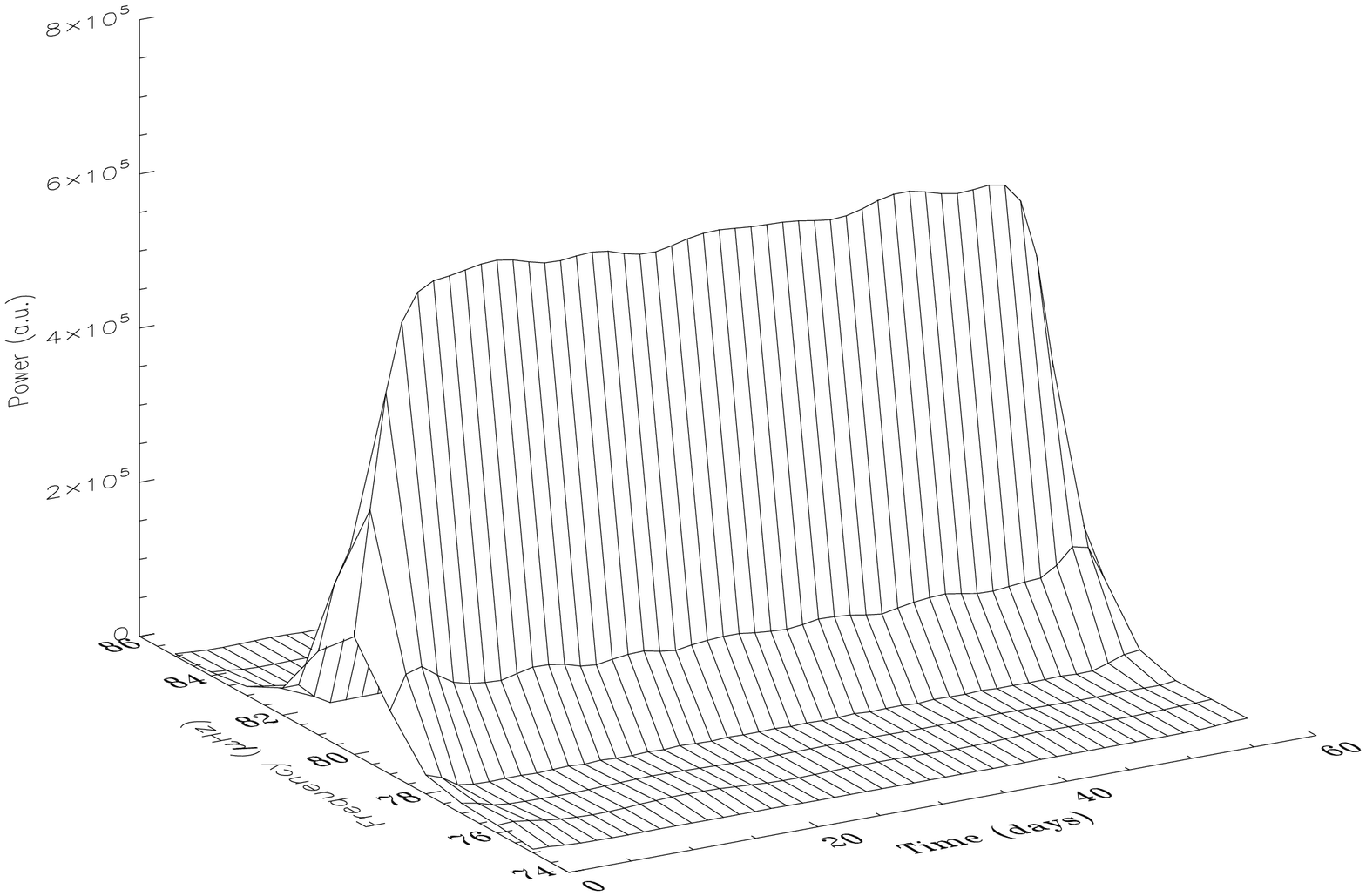}
\includegraphics[angle=90,width=\columnwidth,height=0.75\columnwidth]{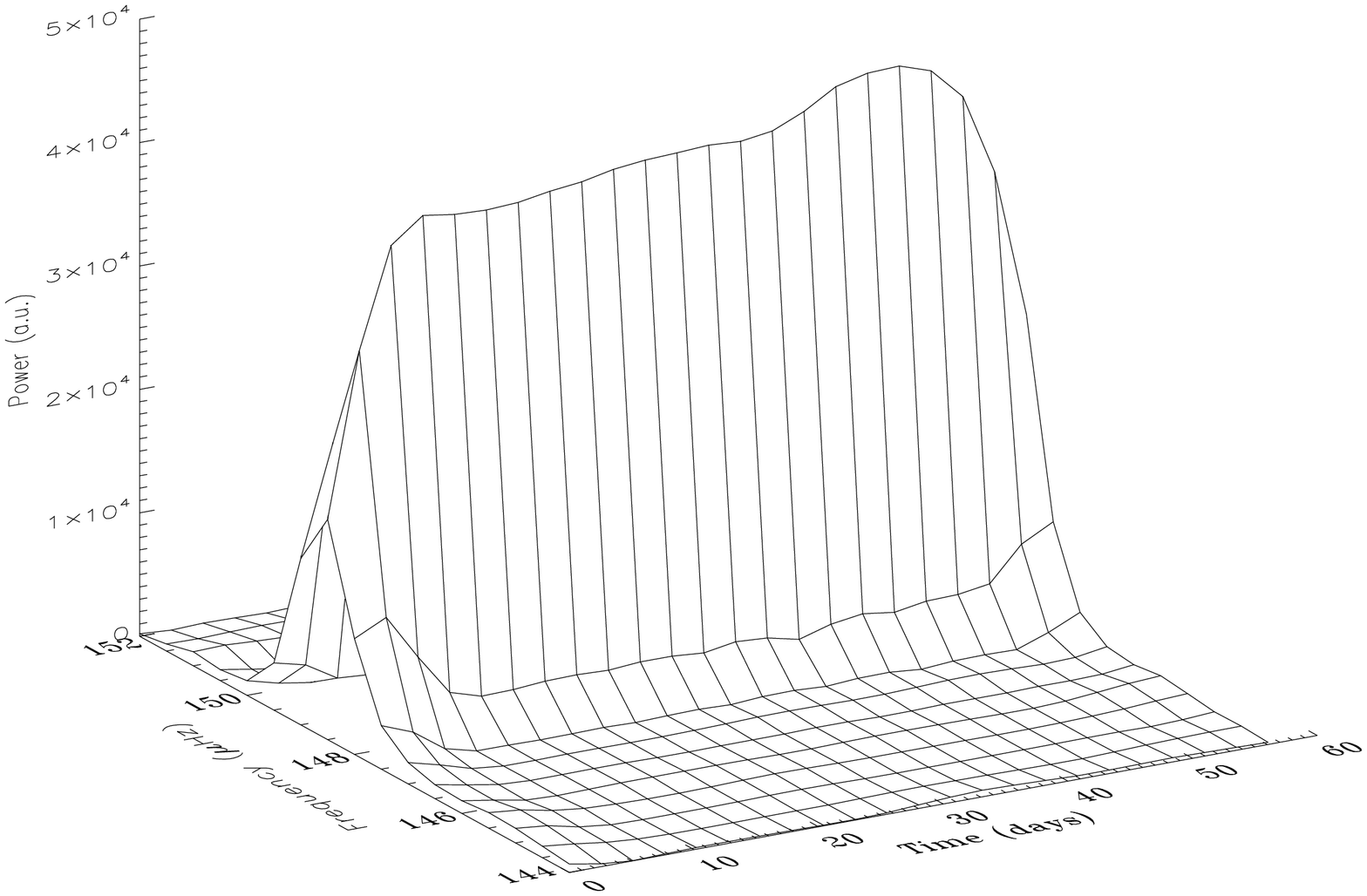}
\caption{Time-frequency analysis of the modes $f_1$=6.92~d$^{-1}$=80.1~$\mu$Hz (top) and 
$f_4$=12.85~d$^{-1}$=148.7~$\mu$Hz (bottom)
with a frequency resolution of 2 and 1~$\mu$Hz, respectively. }
\label{baudin}
\end{center}
\end{figure}

We also investigated the constancy of the highest amplitude terms, i.e.,
those listed in  Table~\ref{comb}.
We subdivided the CoRoT timeseries into  five subsets, each spanning 11.6~d,
 and we calculated a least--squares fit of each subset. 
The amplitude of the predominant term $f_1$ term is the most interesting
to be investigated, since a low percentage of variability will immediately result
in a detectable effect. The amplitude values  span
a full range of 0.10~mmag, with a formal error on each value  
of $\pm$0.05~mmag. Therefore, the amplitude of the $f_1$ term can be considered constant. 
No significant amplitude variability has been detected in the other terms, except 
for $f_6, f_{11}$, and
$f_{12}$. These terms show  evident amplitude and phase variations. This fact is
a known effect induced by the interference of terms closer than the frequency resolution
\citep{doublet}. 
A second investigation of the temporal behaviour of the amplitudes was performed 
using a time-frequency analysis \citep[described in][]{Baudin94}. This analysis is based 
on a Morlet wavelet with an adjustable frequency (and thus, time) resolution. If this 
resolution is greater than the frequency difference between two neighbouring terms, the power 
variation shows the beating described above. If the
resolution is smaller than the  frequency difference, the amplitude of the terms can be 
evaluated. We did not found any trace of variability of the amplitudes
by performing this test on the terms considered above.
The constant behaviour of the amplitudes is illustrated for the terms $f_1$ and  $f_4$ 
in Fig.~\ref{baudin}: the small variations observed are compatible with the influence of the noise. 

Linear combinations of the terms having the highest amplitudes
are commonly found in multiperiodic radial pulsators (Cepheids, RR Lyr, high amplitude
$\delta$ Sct stars), and are also observed  in low amplitude
$\delta$ Sct stars, e.g., FG Vir \citep{fgvir}. 
In the solution of HD~50844 we find some frequencies with a high rank of
detection that could be
explained as a linear combination of a frequency with a low rank and  
the first, predominant term $f_1$ 
(Table~\ref{comb}). However, we cannot rule out that  these high--rank frequencies
 are actually independent modes, excited by resonance mechanisms or even intrinsically excited.

\begin{table}
\setcounter{table}{2}
\begin{flushleft}
\caption{First 20 frequencies identified in the amplitude spectrum.
}
\begin{tabular}{r rr c}
\hline \hline
\noalign{\smallskip}
\multicolumn{1}{c}{Term} & \multicolumn{1}{c}{Frequency}&
\multicolumn{1}{c}{Ampl.} & Possible combination  \\
\multicolumn{1}{c}{}& \multicolumn{1}{c}{[d$^{-1}$]} & 
\multicolumn{1}{c}{[mmag]} &terms \\
\noalign{\smallskip}
\hline
\noalign{\smallskip}
 $f_1$    &  6.92528 & 15.4539 & $f_{29}=2f_1$, $f_{215}=3f_1$ \\
 $f_2$    & 11.21669 &  6.7220 & $f_{112}=f_1+f_2$, $f_{123}=f_2-f_1$\\
          &          &         &  $f_{414}=2f_2$ \\
 $f_3$    & 11.25807 &  4.0143 & $f_{159}=f_1+f_3$, $f_{806}=f_3-f_1$ \\
 $f_4$    & 12.84846 &  3.3603 & $f_{129}=f_1+f_4$, $f_{791}=f_4-f_1$ \\
 $f_5$    & 12.23831 &  3.3271 & $f_{217}=f_1+f_5$, $f_{895}=f_5-f_1$ \\
 $f_6$    & 14.44689 &  3.1767 & $f_{118}=f_1+f_6$\\
 $f_7$    & 13.27347 &  2.6672 & $f_{356}=f_1+f_7$\\
 $f_8$    & 13.35865 &  2.3077 & $f_{149}=f_1+f_8$, $f_{736}=f_8-f_1$ \\
 $f_9$    & 11.75102 &  1.4372 & \\
 $f_{10}$ &  5.26653 &  1.1277 & $f_{64}=f_1+f_{10}$\\
 $f_{11}$ & 14.46126 &  1.0604 & \\
 $f_{12}$ & 14.43209 &  0.9731 & \\
 $f_{13}$ &  9.95254 &  0.9385 & $f_{781}=f_{13}-f_1$\\
 $f_{14}$ & 11.98380 &  0.8268 & $f_{619}=f_1+f_{14}$\\
 $f_{15}$ &  6.55624 &  0.8259 & $f_{141}=f_1+f_{15}$\\
 $f_{16}$ &  7.40446 &  0.8099 &  \\
 $f_{17}$ & 13.56880 &  0.7768 & $f_{497}=f_{17}-f_1$\\
 $f_{18}$ & 10.26062 &  0.7332 & $f_{309}=f_1+f_{18}$\\
 $f_{19}$ &  5.78177 &  0.7046 &  \\
 $f_{20}$ &  6.62914 &  0.6435 & $f_{494}=f_1+f_{20}$\\
\noalign{\smallskip}
\hline
\label{comb}
\end{tabular}
\end{flushleft}
\end{table}

\begin{table}
\begin{flushleft}
\caption{The frequencies identified in high--resolution
spectroscopy. 
}
\begin{tabular}{r rrr c cc}
\hline \hline
\noalign{\smallskip}
\multicolumn{1}{c}{Term} & \multicolumn{1}{c}{Frequency}&
\multicolumn{1}{c}{Spectr.} & \multicolumn{1}{c}{SNR} &
\multicolumn{1}{c}{$\ell,m$} &
\multicolumn{1}{c}{EW} & \multicolumn{1}{c}{RV}  \\
\multicolumn{1}{c}{}&  \multicolumn{1}{c}{[d$^{-1}$]} &
\multicolumn{1}{c}{Ampl.} & \multicolumn{1}{c}{} &
\multicolumn{1}{c}{} & \multicolumn{1}{c}{[km s$^{-1}$]} & \multicolumn{1}{c}{[km s$^{-1}$]} \\
\noalign{\smallskip}
\hline
\noalign{\smallskip}
$f_{12}$   & 14.43209 & 0.27 & 28.6 & 4,2  &       & 0.10 \\
$f_{22}$   & 14.47568 & 0.24 & 22.2 & 3,1  &  \\
$f_{6}$    & 14.44689 & 0.24 & 23.0 & 4,3  & 0.009 & \\
$f_{7}$    & 13.27347 & 0.18 & 15.8 & 3,2  & 0.009 & \\
$f_{3}$    & 11.25807 & 0.15 & 13.1 & 5,1  &  \\
$f_{4}$    & 12.84846 & 0.14 & 19.5 & 3,3  & 0.008 & 0.52 \\
$f_{243}$  & 15.54114 & 0.14 & 14.6 & 9,2  &  \\
$f_{50}$   & 15.22329 & 0.13 & 10.9 & 12,10 & \\
$f_{18}$   & 10.26062 & 0.13 & 12.5 & 5,0  & \\
$f_{29}$   & 13.85029 & 0.13 & 13.4 & 2$f_1$   &       &      \\
$f_{69}$   & 13.16891 & 0.12 & 13.0 & 8,2  & \\
$f_{452}$  & 16.50451 & 0.12 &  9.2 & 14,14 & \\
$f_{8}$    & 13.35865 & 0.10 &  8.9 & 5,3  &       & 0.11 \\
$f_{1}$    &  6.92528 & 0.11 &  5.0 & 0,0  & 0.061 & 0.72 \\
$f_{2}$    & 11.21669 & 0.11 &  9.5 & 3,1  & 0.022 & 0.39 \\
$f_{51}$   & 11.64327 & 0,10 &  9.3 & 8,5  &       & 0.08 \\
$f_{521}$  & 15.12491 & 0.10 &  8.1 & 7,4  & \\
$f_{10}$   &  5.26653 & 0.09 &  4.2 & 2,--2 &       & 0.08 \\
$f_{77}$   & 15.80497 & 0.08 &  6.2 & 4,3  & \\
$f_{9}$    & 11.75102 & 0.08 &  6.4 & 11,7  &       & 0.05 \\
$f_{81}$   & 19.71529 & 0.07 &  4.2 & 12,8  & \\
$f_{84}$   &  9.61191 & 0.06 &  4.1 & 10,4  & \\
$f_{27}$   &  5.67472 & 0.06 &  3.1 & 4,--2 & \\
$f_{71}$   & 14.99570 & 0.06 &  6.7 & 6,3  & \\
$f_{46}$   & 19.75025 & 0.06 &  4.1 & 14,12 & \\
$f_{43}$   &  5.04299 & 0.05 &  3.5 & 4,--2 &       & 0.12 \\
$f_{32}$   &  5.49058 & 0.05 &  3.4 & 5,--2 & \\
$f_{44}$   & 14.60061 & 0.05 &  4.1 & 6,4  & \\
$f_{5}$    & 12.23831 &   -  &   -  & -   & 0.018 &  0.45 \\
$f_{11}$   & 14.46126 &   -  &  -   & -   &       & 0.20 \\
$f_{13}$   &  9.95254 &   -  &  -   & -   &      & 0.10 \\
$f_{30}$   &  5.41862 &  -   & -    & -   &      & 0.08 \\
$f_{19}$   &  5.78177 &   -  &  -   & -   &      & 0.08 \\
\noalign{\smallskip}
Error      &          & 0.02 &      &     &0.002 & 0.02  \\  
\noalign{\smallskip}
\hline
\label{spectro}
\end{tabular}
\end{flushleft}
\end{table}

\section{Ground--based observations}
\subsection{High--resolution spectroscopy:  line profile variations}
The spectroscopic observations were completed from  January 2 to
28, 2007 
with the FEROS instrument mounted at the {2.2-m} ESO/MPI telescope,
La Silla, Chile.
The spectral resolution of FEROS is $R\sim 48,000$. We obtained 232 spectra
in 14  nights. The exposure time was set to 900~sec and
the signal-to-noise ratios (SNRs) ranged from 150 to 210. The spectra were reduced
using an improved version of the standard FEROS pipeline, written in MIDAS and
developed by \citet{monica}.
In addition to the individual line profiles, we considered the mean profiles
computed using the least--squares deconvolution (LSD) method \citep{lsd}. 
When calculating the LSD profiles, the spectral region between 4150 and 5800 $\AA$ was used,
 taking care to omit the 
intervals  containing the $H_{\beta}$ and $H_{\gamma}$ lines. 
By calculating a single line profile with an average SNR of about 1200 for each spectrum, 
we increased the original SNR of the individual spectra by a factor of six.
The mean barycentric radial velocity of the star is $-10.8\pm$0.2~km~s$^{-1}$.
We also derived  a $v\sin~i$ value of 58$\pm$2~km~s$^{-1}$
from {\it i)} the first zero--point position of the Fourier transform for the mean
LSD profile and
{\it ii)} a nonlinear least-squares fit of an intrinsic profile computed 
for the stellar physical parameters of HD~50844 convolved with a rotationally broadened one.
 
Before presenting the results of the frequency analysis, we 
clearly note that the shorter baseline of the spectroscopic
observations (27~d) implies a frequency resolution worse than that of the
CoRoT photometric data. 
The power spectrum of the line profile variations (LPVs) detected in 
the spectroscopic data is shown in  the bottom panel of Fig.~\ref{win}. Since 
the observations are
single--site, the spectral window is  more complicated because of the
presence of relevant aliases at integer values of ~d$^{-1}$ from the
real peak. These facts result in uncertainties on
the amplitudes and on the reliable detection of the spectroscopic
frequencies.  Nevertheless, the study of the LPVs
is of paramount importance for corroborating and complementing the photometric  results,
as suggested by the different appearances of the photometric and
spectroscopic power spectra (Fig.~\ref{win}, middle and bottom panels).

The spectroscopic frequencies were searched for by analysing the LPVs
by means of the pixel--by--pixel technique \citep{manvie,zima}.
Table~\ref{spectro} lists the terms detected in
order of decreasing spectroscopic amplitude \citep[in continuum units as defined by][]{zima},
which is a measurement of the contribution of the given term to the 
line profile variability.
We detected 27 terms with an SNR higher than 4.0 and 3 terms with 3.1$<$SNR$<$3.5
(see Table~\ref{spectro}).
The three modes with 3.1$<$SNR$<$3.5 were also considered in the LPV analysis since they
are associated with well--defined CoRoT frequencies.
The spectroscopic frequencies were associated with the closest frequency detected in 
all the photometric solutions.
In the few cases in doubt (4 out of 30), the term with the largest amplitude was preferred.
This assumption did not modify the subsequent $(l,m)$ identification since the amplitude and
the phase diagrams did not appreciably change with the frequency used to fold the spectroscopic
data. A different 
association with a photometric term would have modified the photometric amplitude, but this parameter
cannot be used in the spectroscopic mode identification.

\begin{figure}[]
\begin{center}
\includegraphics[width=\columnwidth,height=\columnwidth]{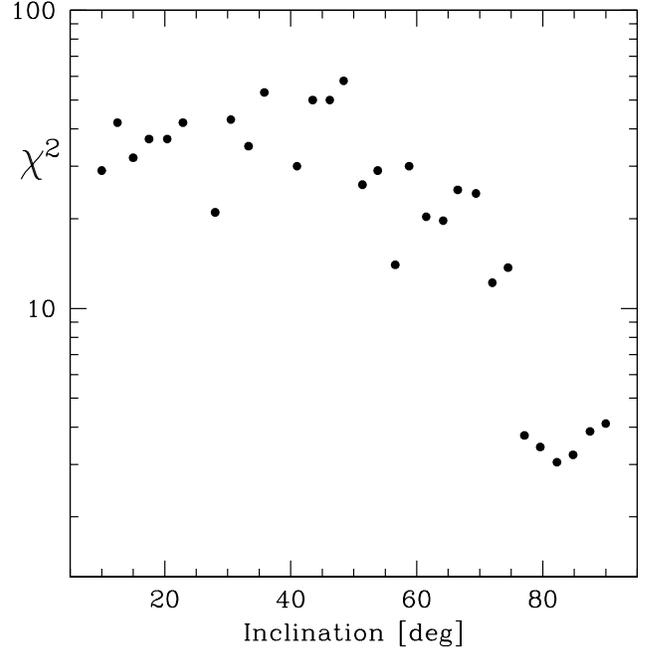}
\caption{\footnotesize 
Behaviour of the $\chi^2$ values, obtained by fitting all the 30
spectroscopic modes to the LSD profiles, each time keeping
the same fixed value for the inclination angle. 
}
\label{incli}
\end{center}
\end{figure}

The mode identification 
was performed by fitting the amplitude and phase variations of each mode across the line profile
\citep{manvie} by using the software FAMIAS \citep{famias}.
The identified ($\ell$,$m$) values are listed in Table~\ref{spectro}. 
In the ($\ell,m$) couples, the negative $m$ values indicate retrograde modes.
Uncertainties 
are estimated to be  $\pm$1 for the degree  $\ell$ and $\pm$2 for the order  $m$. 
They are larger in the cases of the modes $f_{12}$ and, in particular,  $f_{22}$ and $f_6$, since 
they are very close in frequency. Also the nonsignificant detection of $f_{11}$ is probably a 
problem related to the low resolution in the spectroscopic data. 
Mainly tesseral modes were detected, but also a couple of sectoral modes were found.
This technique also returned a reliable estimate of the inclination
angle, i.e.,  $i$=82$\pm$4~deg (Fig.~\ref{incli}). Therefore, HD~50844 is seen almost equator--on.  The associated 
equatorial rotational velocity is not significantly different from the $v\sin i$ value 
derived above.
This particular orientation explains why only one axisymmetric mode (i.e.,
$f_{18}$, Table~\ref{spectro}) was observed: these modes produce marginal variations
along our line of sight.
                         
An independent search for frequencies was performed on the timeseries defined by the $0^{\rm th}$ 
(equivalent width, EW) and $1^{\rm st}$ (radial velocity, $\langle v\rangle$) order moments. 
Since these quantities are integrated over the whole
stellar disk, they are more sensitive to reveal low--degree modes. 
The amplitudes of the frequencies, detected in both moments by  the Period04 \citep{period} method, 
are listed in Table~\ref{spectro}.
We recover  in the EW and $\langle v\rangle$ variations  (bottom of Table~\ref{spectro}) 
some of the terms with a large photometric amplitude that have been missed  in the LPV analysis, 
because of their  intrinsic  low spectroscopic amplitude. 
 Since they were detected in the first moments, we can speculate that it probably concerns low--degree modes.

The six frequencies detected in the EW variations correspond to six out of the  
seven first photometric frequencies. The term $f_3$ is probably missing because it is  
not well-resolved from $f_2$ and  $f_5-1$~d$^{-1}$.
The correlation between the terms found in the $1^{\rm st}$ moment and those detected in
the CoRoT photometric data is not as clear as in the case of the $0^{\rm th}$ moment.
We note  that the frequency 13.85029~d$^{-1}$=2$f_1$  was  detected 
in the pixel-to-pixel variations (Table~\ref{spectro}). Does it really concern the first harmonic of $f_1$
 or is it an independent mode excited by resonance?
We tested the hypothesis following the method described by \citet{telting}, which states 
that a frequency is a first harmonic if the phase difference $\psi_{01}=2\psi_0-\psi_1$ 
between the line-centre phases of the two frequencies is 1.50~$\pi$~rad. Because we  
measure $\psi_{01}=1.58~\pi\pm0.19$~rad, we conclude that  13.85029~d$^{-1}$ is indeed 
the first harmonic of $f_1$.

\begin{figure}[]
\begin{center}
\includegraphics[width=\columnwidth,height=\columnwidth]{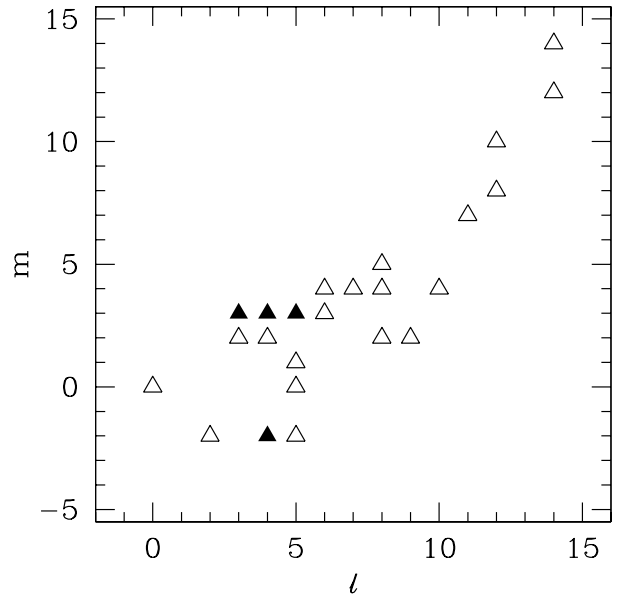}
\caption{\footnotesize Identification of the ($\ell,m$) couples from high--resolution
spectroscopy. 
Open triangles: one term with that ($\ell,m$)  identification.
Filled triangles: two terms with that ($\ell,m$) identification.
 }
\label{lm}
\end{center}
\end{figure}

The main goal of the spectroscopic observations
is to answer the crucial question about the huge number of
 frequencies detected in the CoRoT photometric data: 
is the large number of photometric terms related to the visibility not only of low--degree modes but
also of high--degree modes? 
To investigate this we focus on the mode identification results, as derived from the high-resolution data, 
presented in Table~\ref{spectro}. In  Fig.~\ref{lm} we plot the identified degree $\ell$  
in function of order $m$.
This figure shows  that  the high--degree modes also tend to have  a high order $m$.
The highest detected degree is  $\ell=14$. This observational fact tells us that
we have at least 235 possible modes of pulsation for a given radial order $n$.
 A few  different radial orders are
sufficient for explaining the 1000 frequencies detected in the CoRoT photometric
timeseries of HD~50844. 
On the other hand, 
we know from CoRoT photometry that the frequency values higher than 30~d$^{-1}$
are observed after the 300$^{\rm th}$
rank of detection, i.e.,  with very low amplitudes. The observed spectroscopic
frequencies have $f<$20~d$^{-1}$.  We still miss the spectroscopic
counterpart of the {$20<f<50$~d$^{-1}$} region. Such a region could be
filled by  modes with high radial orders $n$, which are shifted
toward higher frequencies than those of the low orders.

The excitation of the almost totality of the modes allowed within a 
narrow radial order interval
and its possible extension to a wide range of radial orders  
can explain the presence of 1000 terms in quite a natural way.
In either case, we can state that the huge number of photometric
terms are plausible since  the spectroscopic data are confirming that
we are observing modes with high ($\ell, m$) values. In turn, this means that the cancellation
effects for modes up to  $\ell\sim14$ are not working and that these modes are able to 
produce a net flux measurable at the 10$^{-5}$~mag level.


\subsection{High--resolution spectroscopy: abundance analysis}
In addition to the LPV analysis, the
FEROS spectra  were used to determine the  abundances of the
elements.  The atmospheric models were
computed with the line-blanketed LTE ATLAS9 code \citep{kurucz}, which
handles line opacity with the opacity distribution function. The
synthetic spectra were computed with the SYNTHE code \citep{kurucz}. 
The stellar-line identification and
the abundance analysis were performed on the basis of the VALD lines
list \citep{kupka}. 
The mean LTE abundances $\log\epsilon(\rm{El})$ (by convention,
$\log \epsilon$[H] = 12), along with the uncertainties
determined as $\sigma = \sqrt{\sigma_{\rm T}^2+ \sigma_{\rm \log g}^2}$
are reported in Table~\ref{ewa}. The atmospheric parameters 
$T_{\rm eff}$ = 7400\,K, $\log g$ = 3.6 and $\xi$ = 2\,{km~s}$^{-1}$ 
were adopted. To
determine the uncertainties connected with effective temperature and
surface gravity, we calculated the abundances for atmospheric models with
$T_{\rm eff} = \pm$\,200\,K and $\log g = \pm$ 0.2. 

\begin{figure}[]
\begin{center}
\includegraphics[width=\columnwidth,height=0.5\columnwidth]{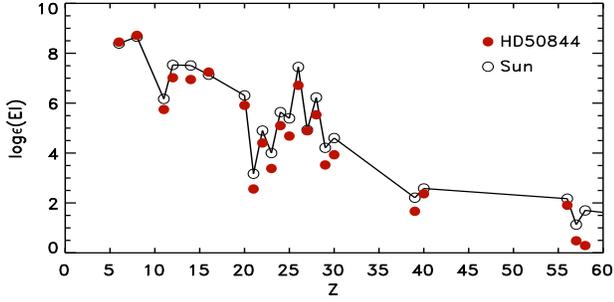}
\caption{\footnotesize Element abundances of HD~50844 derived from FEROS
spectra compared to the solar values \citep{grevesse}.
}
\label{ewaplot}
\end{center}
\end{figure}

As a general result, HD~50844 is a slightly metal-deficient star. However,
Table~\ref{ewa} shows how the abundances of elements as C, N, and S
in the HD~50844 atmosphere are very similar to those of the Sun, while
other elements are underabundant (see also Fig.~\ref{ewaplot}).  Such an 
abundance pattern is  typical of $\lambda$ Boo stars \citep{paunzen}. With respect to the
claimed spectroscopic binarity of these stars \citep{gerbaldi}, we 
did not find any evidence  of a second component  in the FEROS spectra, 
and we did not detect a long--term radial velocity drift that could be ascribed to an orbital
motion. 

\begin{table}
\begin{flushleft}
\caption{Mean LTE abundances $\log\epsilon(\rm{El})$ 
along with the uncertainties, $\sigma$, with
 solar abundances determined by \citet{grevesse}.}
\begin{tabular}{lcccc}
\hline
\hline 
\noalign{\smallskip}
\multicolumn{1}{c}{Element} & \multicolumn{1}{c}{Z}&
\multicolumn{1}{c}{$\log\epsilon(\rm{El})$} & \multicolumn{1}{c}{$\sigma$}&  Sun \\
\noalign{\smallskip}
\hline
\noalign{\smallskip}
 C & 6  &  8.45 & 0.130 &  8.39 \\
 N & 8  &  8.72 & 0.102 &  8.66 \\
 Na& 11 &  5.75 & 0.097 &  6.17 \\
 Mg& 12 &  7.02 & 0.113 &  7.53 \\
 Si& 14 &  6.95 & 0.386 &  7.51 \\
 S & 16 &  7.25 & 0.054 &  7.14 \\
 Ca& 20 &  5.91 & 0.112 &  6.31 \\
 Sc& 21 &  2.56 & 0.103 &  3.17 \\
 Ti& 22 &  4.40 & 0.159 &  4.90 \\
 V & 23 &  3.38 & 0.146 &  4.00 \\
 Cr& 24 &  5.10 & 0.140 &  5.64 \\
 Mn& 25 &  4.68 & 0.157 &  5.39 \\
 Fe& 26 &  6.71 & 0.126 &  7.45 \\
 Co& 27 &  4.89 & 0.098 &  4.92 \\
 Ni& 28 &  5.53 & 0.151 &  6.23 \\
 Cu& 29 &  3.52 & 0.018 &  4.21 \\
 Zn& 30 &  3.93 & 0.079 &  4.60 \\
 Y & 39 &  1.67 & 0.099 &  2.21 \\
 Zr& 40 &  2.37 & 0.458 &  2.58 \\
 Ba& 56 &  1.91 & 0.170 &  2.17 \\
 La& 57 &  0.483& 0.126 &  1.13 \\
 Ce& 58 &  0.292& 0.436 &  1.70 \\
\noalign{\smallskip}
\hline
\label{ewa}
\end{tabular}
\end{flushleft}
\end{table}

\subsection{Multicolour photometry}
The importance of colour information to estimate the degree $\ell$
of the modes is   well-known  \citep{garvie}; therefore, 
we observed HD~50844 in Stromgren $uvby$ photometry
to accompany the CoRoT white--light photometry. 
Several campaigns were carried out 
at S. Pedro M\'artir and Sierra Nevada observatories, using  twin
Danish photometers. The multicolour observations span 743~d. 
 The campaigns  started when the IR01 pointing was decided and continued
after the CoRoT observations. The final $uvby$ dataset
consists of 1496  datapoints, collected in 66 nights for
a total survey of 317.4~hours.
Unfortunately, the space and ground-based observing windows are not 
simultaneous.
The spectral window of the ground -based data is characterized by
several alias peaks (at integer values of d$^{-1}$ and ys$^{-1}$ of the central peak)
and  the true peaks could only be identified  knowing the CoRoT
solution.
We considered the first 20 CoRoT frequencies and refined their values by
performing a least--squares fit of the data in the $v$ colour, i.e., the one
with the largest amplitude. This step was necessary since the
time baseline of the $uvby$ data is much longer than that of the CoRoT
data. The first 13~frequencies resulted in a significant $v$-amplitude.
The $v-$amplitude of the $f_1$ term is 22.2$\pm$0.3 mmag, while  
those of the others are much lower (e.g., 8.0~mmag for $f_2$ and
1.70~mmag for $f_{13}$). 

\begin{figure}[]
\begin{center}
\includegraphics[width=\columnwidth,height=0.6\columnwidth]{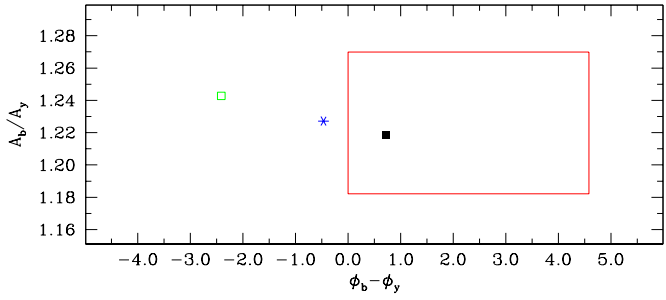}
\caption{\footnotesize
Comparison between theoretical predictions and multicolour photometric observations for $f_1$, the
mode with the largest photometric amplitude. Amplitude ratios vs.  phase differences (in degrees)
in $b$ and $y$ colours are shown.  
Red large square is the observed errorbox.
Points represent theoretical predictions for spherical degree $\ell$=0
(black filled square), $\ell$=1 (blue star), and $\ell$=2 (green open square).
}
\label{andy}
\end{center}
\end{figure}

We calculated the amplitude ratios and the phase shifts  by using
the $y$ colour as the reference system \citep{garvie}. They were compared with the
theoretical predictions computed using the non-adiabatic pulsational code GraCo
\citep{graco}.
The immediate result is that the predominant term $f_1$=6.92~d$^{-1}$ 
is characterized by positive phase shifts (i.e., $\phi_{u,v,b}-\phi_y>0.0$).
This  puts a very tight constraint on the mode
identification, since only the radial ($\ell$=0) mode  shows positive phase shifts in the    
$\phi_b-\phi_y$ case (Fig.~\ref{andy}).
On the other hand, the comparison between the observed and the calculated
values does not supply conclusive mode identifications in the cases of the other
terms. 

To further investigate the $f_1$ mode identification, we note that the harmonics 2$f_1$=$f_{29}$
(amplitude 0.56~mmag) and 3$f_1$=$f_{215}$ (0.12~mmag) have been
detected in the CoRoT data. 
 The observed values differ from the exact multiples of $f_1$ by $3~\times~10^{-4}$ and  
$1~\times~10^{-3}$~d$^{-1}$, respectively.
 Both the $\phi_{21}$=4.04~rad and the $\phi_{31}$=2.11~rad 
Fourier parameters  are very similar to those observed in high amplitude
$\delta$ Sct stars \citep[Figs.~4 and 5 in][]{ogle},
known to pulsate in the fundamental (rarely in the first overtone) radial mode. 
Identification of $f_1$ as a radial mode
is also consistent with being the highest amplitude term
 both in radial velocity and in equivalent width. 
Although a nonradial mode could display such large variations,
such variations are usually connected with the large displacements typical of 
radial modes.
The mode identification of the $f_1$ term from the spectroscopic data was not trivial since we
had to take the harmonic 2$f_1$ and the EW variations into account. 
The FAMIAS method returns two possibile $(\ell,m)$ couples, i.e., 
(0,0) and (2,0). When considering both the spectroscopic and the photometric identifications,
we can consider the $f_1$ term as the fundamental radial mode. 

\section{The seismic inferences}
The discussion of  the number of excited frequencies  in $\delta$ Sct stars
has followed a particular pattern. At the beginning of the observational efforts
to discover as many modes as possible \citep{porvie,brevie}, the number of
detected frequencies was less than expected and a kind of selection
effect in exciting modes/damping amplitudes was claimed \citep[][
and, more recently, \citealt{lenz}]{guzik}.
\citet{nz} suggest that the problem was mainly of an  SNR nature. They stressed
how the number of detected terms in the science case of FG Vir follows an exponential behaviour 
as a function of the observational technique used, namely from single--site observations 
in a few nights to multisite observations in several years. In a certain sense,  
they correctly predicted that a long--term, high--precision space 
mission as  CoRoT would allow access to the distribution of the amplitudes
down to $10^{-5}-10^{-6}$~mag, allowing the harvest of a multitude of modes.
That the distribution of the frequencies is quite uniform in the 0--20~d$^{-1}$
region (Fig.~\ref{isto}) implies that modes are excited at lower frequencies
than that of the fundamental radial mode, i.e., $f<$6.9~d$^{-1}$.
The peaks with $f<3~$d$^{-1}$ can be enhanced by 
an incomplete removal of the instrumental drift.
The spectrum of HD 292790 (Fig.~\ref{spectrum}, top panel) shows a noise level
below 0.05 mmag in this region and therefore this contamination should
become important only after the 600$^{\rm th}$ frequency (see Table~\ref{lsq}).
We can argue from the CoRoT timeseries on HD~50844 that
gravity and mixed modes are quite common in $\delta$ Sct stars.

The frequency set obtained from the CoRoT timeseries of HD~50844 allows us to investigate the
problem of the regular spacings in the frequency spectrum of 
$\delta$ Sct stars. There have been several claims that such spacings will be 
modified by nonlinear effects and mixed modes, especially in evolved stars
\citep{jcd,goupil,guzik}.
This hint is confirmed by the frequency spectrum of HD~50844,
which does not supply any reliable indication of the regular spacings
detected in solar--like pulsators \citep[e.g.,][]{science}.

\begin{figure}
\includegraphics[width=\columnwidth,height=\columnwidth]{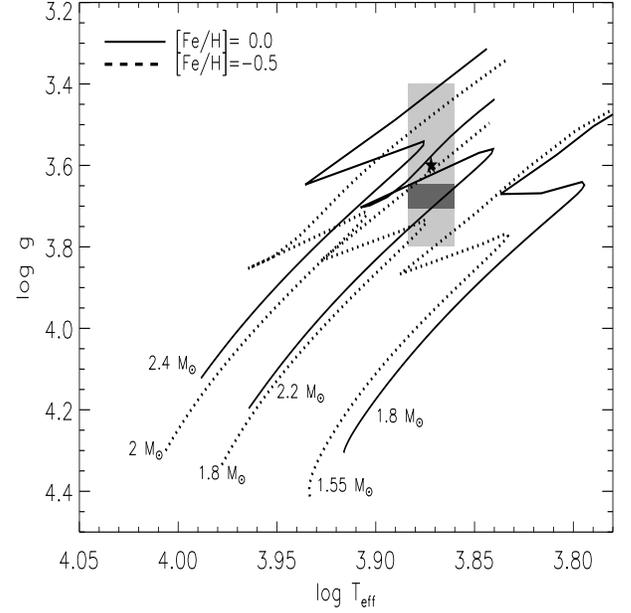}
     \caption{HR diagram containing several evolutionary tracks of
              representative models of the star. The shaded area represents
              the observational uncertainties in effective temperature
              and gravity. The models in the darker area have
              a fundamental radial mode equal to $f_1$=6.925~d$^{-1}$. 
              The filled star represents the observational
              average location of HD~50844 in the diagram. Continuous
              lines represent tracks of  evolved models with a solar
              metallic content, whereas dotted lines represent evolutionary
              tracks of submetallic models; the  latter are more representative
              of the HD~50844 case. 
}
               \label{suarez}
\end{figure}

As a first attempt to model the  scientific case of HD~50844, equilibrium models representative 
of the star have been calculated with the evolutionary code
{\sc cesam}  \citep{Morel97}, taking first-order effects of rotation into
account.
Such models (the so-called pseudo-rotating models) include the spherically
averaged contribution of the centrifugal acceleration by means of an
effective gravity. The nonspherical components of the
centrifugal acceleration (not considered in the equilibrium models)
are included in the adiabatic oscillation computations by means of a linear
perturbation analysis \citep{Sua06rotcel}.
Figure~\ref{suarez} shows the models within the uncertainty box 
in T$_{\rm eff}$ and $\log g$. The models,
evolved considering the local conservation of the angular momentum,
 are able to return a surface rotational
velocity around 60~km~s$^{-1}$ (i.e., the value obtained from $v\sin i$
taking $i=80$~deg as suggested by the analyses of the
the line profile variations) starting from a value of 75~km~s$^{-1}$ on the ZAMS. 
We can also compute  the fundamental radial mode of
our models using the code GraCo \citep{graco}: the darker box in Fig.~\ref{suarez}
marks the cases matching the observed value $f_1$. When considering the
slightly metal deficient content of HD~50844, we can argue that the models suggest a mass 
between 1.65 and 1.80~$M_{\sun}$.
 
\section{Conclusions}
The exploitation of the CoRoT photometric timeseries of HD~50844 resulted in    
a very complex and intriguing task, also giving a completely new picture of the 
pulsational content of $\delta$ Sct stars. We demonstrated that
the light curve of HD~50844 can be explained by the presence of
hundreds of excited terms. Classical checks such as using  different
software packages, subdividing the timeseries into different subsets,
carefully inspecting the residuals, and  comparing with other similar targets
observed by CoRoT, together confirm this issue. High--resolution spectra support
the excitation of modes having a very high degree $\ell$ (up to $\ell$=14), 
thus providing an observational explanation  for the richness of the frequency spectrum.
Such a large number of modes of different $\ell$ simultaneously excited is among plausible
features suggested by theoretical works \citep[e.g.,][]{dzie}.
An immediate conclusion is that the cancellation effects are not sufficient 
in removing the variations of the flux integrated over the whole stellar disk;
\citet{romapol} correctly predicted this extension 
to high degrees of the  modes
extracted from  high--precision photometric timeseries. 
We should  keep in mind that the CoRoT performances and  the continuous 
monitoring allow  detection of amplitudes of about 10$^{-5}$~mag without significant
aliasing effects, which is  a totally new perspective. 

The case of HD~50844 seems to match the theoretical picture
featuring a very large number of excited modes with amplitudes limited by a saturation
mechanism of the $\kappa$ mechanism.
This confirms that the observation of $\delta$ Sct stars with CoRoT will allow us to 
address the longstanding problem of amplitude distribution and the potential
existence of mode selection processes in the $\delta$ Sct instability strip.
We note that HD~50844  probably does not constitute a ``standard" scientific case 
of a $\delta$ Sct star, because it is 
located on the TAMS,  probably belongs to the class of $\lambda$ Boo stars (i.e., 
showing atmospheric particularities), and it is  seen almost equator-on. 
Nevertheless, seismic models 
that were able to reproduce the position of HD~50844 in the HR
diagram and its measured rotational velocity were calculated.

HD~50844 has for the first time disclosed the intrinsic complexity
of  frequency spectra of intermediate--mass stars in the lower part of the
instability strip.
It is quite evident that, after these first CoRoT observations, we must look at $\delta$ Sct 
stars in a completely new way. We have to realize that ground--based observations were 
only able to observe the tip of the iceberg, while the larger part of the modes have
remained undetected.

\begin{acknowledgements}
The FEROS data are being obtained as part of the ESO Large Programme
LP178.D-0361 (PI.: E.~Poretti). The authors wish to thank the anonymous
referee for useful comments.
Mode identification results were
obtained with the software package FAMIAS developed in the framework of the 
FP6 Coordination Action Helio- and 
Asteroseismology (HELAS; http://www.helas-eu.org/). WZ was  
supported by the FP6 European Coordination Action HELAS and by the Research 
Council of the University of Leuven under grant GOA/2008/04.
This work was supported by the Italian
ESS project, contract ASI/INAF I/015/07/0, WP 03170, and by the Hungarian
ESA PECS project No. 98022. KU acknowledges financial support from a
{\it European Community Marie Curie Intra-European Fellowship},
contract number MEIF-CT-2006-024476. PJA acknowledges financial support
from a {\it Ramon y Cajal} contract of the Spanish Ministry of Education
and Science. AM ackowledges financial support from a {\it Juan de la Cierva}
contract of the Spanish Ministry of Science and Technology. 
SMR acknowledges a {\it Retorno de Doctores} contract financed by the Junta de Andaluc\'{\i}a and
Instituto de Astrof\'{\i}sica de Andaluc\'{\i}a for carrying out observing campaigns for CoRoT targets
at Sierra Nevada Observatory.
EN acknowledges financial support of the N~N203~302635 
grant from the MNiSW.

\end{acknowledgements}

\end{document}